\documentclass[a4paper,fleqn]{cas-sc}

\usepackage{graphicx}
\usepackage{booktabs}
\usepackage{gensymb}
\usepackage[authoryear,longnamesfirst]{natbib}
\usepackage{longtable}
\usepackage{lineno}

\begin{document}

\title{Long-Baseline VLF Observations of Solar Flares from Antarctica}
\shorttitle{Long-Baseline VLF Flare Observations}
\shortauthors{K. Kozarev et~al.}

%%%%%%%%%%%%%%%%%%%%%%%%%%%%%%%%%%%%%%%%%%%%%%%%%%%
%% Authors
%
\author[aff1]{Kamen Kozarev}
\ead{kkozarev@astro.bas.bg}
%\cormark[cor1]

\author[aff2]{Peter Petkov}
\author[aff2]{Ivaylo Nachev}
\author[aff3]{Veselka Radeva}
\author[aff1]{Momchil Dechev}
\author[aff1]{Galin Borisov}
\author[aff3]{Anton Atanasov}

%%%%%%%%%%%%%%%%%%%%%%%%%%%%%%%%%%%%%%%%%%%%%%%%%%%
%% Affiliations
%
\affiliation[aff1]{organization={Institute of Astronomy and NAO, Bulgarian Academy of Sciences},
            city={Sofia},
            postcode={1784},
            country={Bulgaria}}

\affiliation[aff2]{organization={Technical University of Sofia},
            city={Sofia},
            postcode={1756},
            country={Bulgaria}}

\affiliation[aff3]{organization={Nikola Vaptsarov Naval Academy},
            city={Varna},
            postcode={9002},
            country={Bulgaria}}

%\linenumbers
\begin{abstract}
We present long-baseline Very Low Frequency (VLF) observations of solar flare-induced ionospheric disturbances obtained at the Bulgarian Polar Astronomical Observatory (St. Kliment Ohridski Base) on Livingston island, Antarctica. Using continuous VLF transmissions at 21.4 kHz (NPM, Hawaii) and 24.0 kHz (NAA, Maine), propagating over trans-hemispheric paths exceeding 11,000 km, we analyze observations of solar flares during the period 24 January–8 February 2025. After removing the strong diurnal signal via superposed epoch analysis, we analyse the flare-related perturbations in VLF amplitude and their correlation with GOES soft X-ray flux for 250 flares of C and M class. The long propagation paths provide enhanced sensitivity to flare-driven changes in D-region ionization. The observations reveal clear, frequency-dependent responses and measurable time delays between X-ray peaks and VLF extrema. These delays, including cases of near-zero or negative correlation lag for stronger events, are consistent with a possible role for flare spectral characteristics and D-region recombination processes, although alternative explanations related to waveguide propagation and signal-profile differences cannot presently be excluded. Our results demonstrate the scientific value of long-baseline Antarctic VLF observations for detecting and timing GOES-class flares, while also highlighting key limitations, such as background variability and path-dependent propagation effects, which must be quantified for reliable VLF-based flare monitoring.
\end{abstract}

\maketitle

\section{Introduction}
Solar flares cause sudden and significant enhancements in X-ray and extreme ultraviolet (EUV) radiation, which dramatically increase the ionization of the Earth’s atmosphere \citep{Hayes2021Solar, Hayes2017Pulsations}. This phenomenon, known as a Sudden Ionospheric Disturbance (SID), particularly impacts the lowest layer, the D-region (approximately 60–90 km in altitude) \citep{Mitra1974Ionospheric, Whitten1965}. The D-region is critical for space weather because it is primarily responsible for the absorption of High-Frequency (HF) radio waves used for civil aviation, broadcasting, and military communications, often resulting in short-wave fadeouts \citep{Cannon2013, Knipp2016}. Since the D-region is too high for balloons and too low for satellites, and exhibits electron densities too low for standard radar techniques, direct measurement is challenging \citep{Mitra1974Ionospheric}. Therefore, Very Low Frequency (VLF; 3–30 kHz) radio waves provide a unique and effective remote-sensing tool to probe this dynamic layer \citep{Wait1964, Mitra1974Ionospheric}.

VLF waves propagate in the Earth-ionosphere waveguide, where the D-region acts as the upper, reflective boundary \citep{Wait1964}. During a flare, increased X-ray flux (wavelengths $<$~10~\AA) penetrates to D-region altitudes, enhancing ionization and altering the waveguide properties, which manifests as measurable variations in VLF signal amplitude and phase \citep{Thomson2005, McRae2004}. A substantial body of work over the past two decades has established the quantitative relationship between VLF perturbations and solar X-ray flux, effectively allowing the Earth’s ionosphere to act as a natural detector of solar flares \citep{Thomson2004, Todoroki2007}. The resulting electron density enhancements can reach an order of magnitude over short timescales \citep{Hayes2017Pulsations}. Furthermore, VLF records exhibit a characteristic time delay ($\Delta t$), referred to as `sluggishness' \citep{Appleton1953Note} or `relaxation time' \citep{Mitra1974Ionospheric}, between the peak X-ray flux and the peak VLF response. This delay reflects the complex balance between ionization and recombination processes in the D-region \citep{Zigman2007Dregion, Grubor2008, Kolarski2014}.

Recent studies utilizing VLF measurements have revealed new insights while highlighting remaining challenges. Observations confirm that the D-region is sensitive to small-scale variations in soft X-ray flux, exhibiting synchronized pulsations that mirror quasi-periodic pulsations (QPP) observed in flare emissions \citep{Hayes2017Pulsations, Hayes2021Solar}. Investigations of strong flares have further demonstrated that the peak ionospheric response may precede the soft X-ray maximum \citep{Briand2022Role, Hayes2021Solar}. This behavior suggests that higher-energy photons, such as hard X-rays (HXR, $\gtrsim$40~keV), contribute significantly to early D-region ionization \citep{Briand2022Role}, emphasizing the importance of full spectral characterization of flare emissions. Additional observational studies across different propagation paths and latitudes, including polar and long-baseline measurements, further demonstrate the sensitivity of VLF signals to geophysical and solar conditions \citep{Macotela2017}.

In this paper, we present the first VLF flare observations obtained at the St. Kliment Ohridski polar base (`Ohridski base' hereafter) on Livingston Island, Antarctica. The base is located at $-62^\circ38'29''$ latitude and $60^\circ21'53''$ longitude and is operated by the Bulgarian National Center for Polar Research. We analyze the correspondence between two-frequency VLF time series and GOES-18 X-ray observations to evaluate the feasibility and limitations of long-baseline VLF propagation as a reliable proxy for solar flare occurrence and timing. The objective of this work is to assess the robustness of very long-path VLF observations for large-scale flare detection under polar conditions. The paper is structured as follows: Section~\ref{methods} describes the experimental setup and data processing, Section~\ref{results} presents the analysis and results, Section~\ref{discussion} discusses their implications, and Section~\ref{conclusions} summarizes the main findings.

\section{Methods}
\label{methods}

\subsection{Observing setup}
The observing system detects reflected waves from above-ground sources of a continuous signal in the kilohertz range, and the reflections of this signal form an ionospheric-ground waveguide, within the limits of which the signal changes characterize the influence of solar flares on the ionosphere. We observed the response of the ionosphere's D-layer to man-made VLF radio emission at two distinct frequencies. The 21.4~kHz signal is emitted from the United States Navy transmitter (call sign NPM) at Lualualei, Hawaii, USA, with a shortest path length to the Bulgarian Antarctic base of 12 495 km. The 24.0~kHz signal is emitted from the United States Navy transmitter (call sign NAA) at Cutler, Maine, USA, with a shortest path length of 11 909 km.

To detect the VLF emission, a precision VLF observing system was constructed at the Technical University of Sofia to measure amplitude perturbations in received signals. The receiver employs a square magnetic loop antenna consisting of 50 turns of 0.5 mm copper wire wound on a 1.15 m side frame (area 1.32 m²), providing an inductance of approximately 7.2 mH. The loop is tuned with an approximately 8 nF polypropylene capacitor to form a parallel resonant circuit. The resonant response of the complete receiving front end was experimentally characterized by magnetically injecting a pilot signal into the loop and monitoring the corresponding response at the output of the amplification chain. The measured resonant frequency was approximately 22 kHz, with an unloaded quality factor of approximately 65. After connection to the input matching network, the loaded quality factor was approximately 10.7, corresponding to a nominal 3-dB bandwidth of approximately 2 kHz. Although the 24.0 kHz signal lies outside the nominal 3-dB bandwidth, sufficient off-resonance sensitivity is retained for reliable reception. The frequency-dependent roll-off of the resonant circuit is static and therefore introduces only a fixed frequency-dependent scaling at the two monitored frequencies, without affecting their temporal variations.
 
 The amplification chain uses a balanced-input low-noise instrumentation amplifier followed by two NE5532 gain stages, providing an overall voltage gain of approximately 200. The input amplifier has a specified maximum input voltage-noise density of 3.2 nV/$\sqrt(Hz)$ and an input current-noise density of 200 fA/$\sqrt(Hz)$ at 1 kHz. The linearity and gain stability of the complete amplification chain were verified experimentally on the laboratory bench prior to deployment. Although the SDR has a relatively high specified noise figure of 19 dB at 300 kHz, its input-referred contribution to the overall system noise is strongly reduced by the approximately 200-fold voltage gain of the preceding low-noise analog front end and is therefore expected to be negligible compared with the front-end contribution.
 
 The amplified VLF band is digitized by the SDR within a single frequency window containing both the 21.4 and 24.0 kHz signals. Both frequencies are therefore acquired simultaneously using the same receiver configuration, analog gain, digitization chain, and time base, without retuning or switching between channels. During subsequent processing, the received power at 21.4 and 24.0 kHz is extracted from the corresponding frequency bins within the common SDR acquisition window and recorded as two independent time series.

%[TO DO:]\\
%- Correlation between the overall 21.4 and 24.0 kHz signals\\
%- Make plots of the flare correlation coefficients vs. flare class\\
%- Check for negative onset delays (VLF onset first). In those cases, there may be hard X-ray emission which causes the detection in the ionosphere. For those cases, check which frequency was detected better, 21 or 24, might have to do with the 
%- Compute the time delay between the peaks in X-ray and peaks in VLF.\\

\begin{figure*}[t]
\centering
\includegraphics[width=0.8\textwidth]{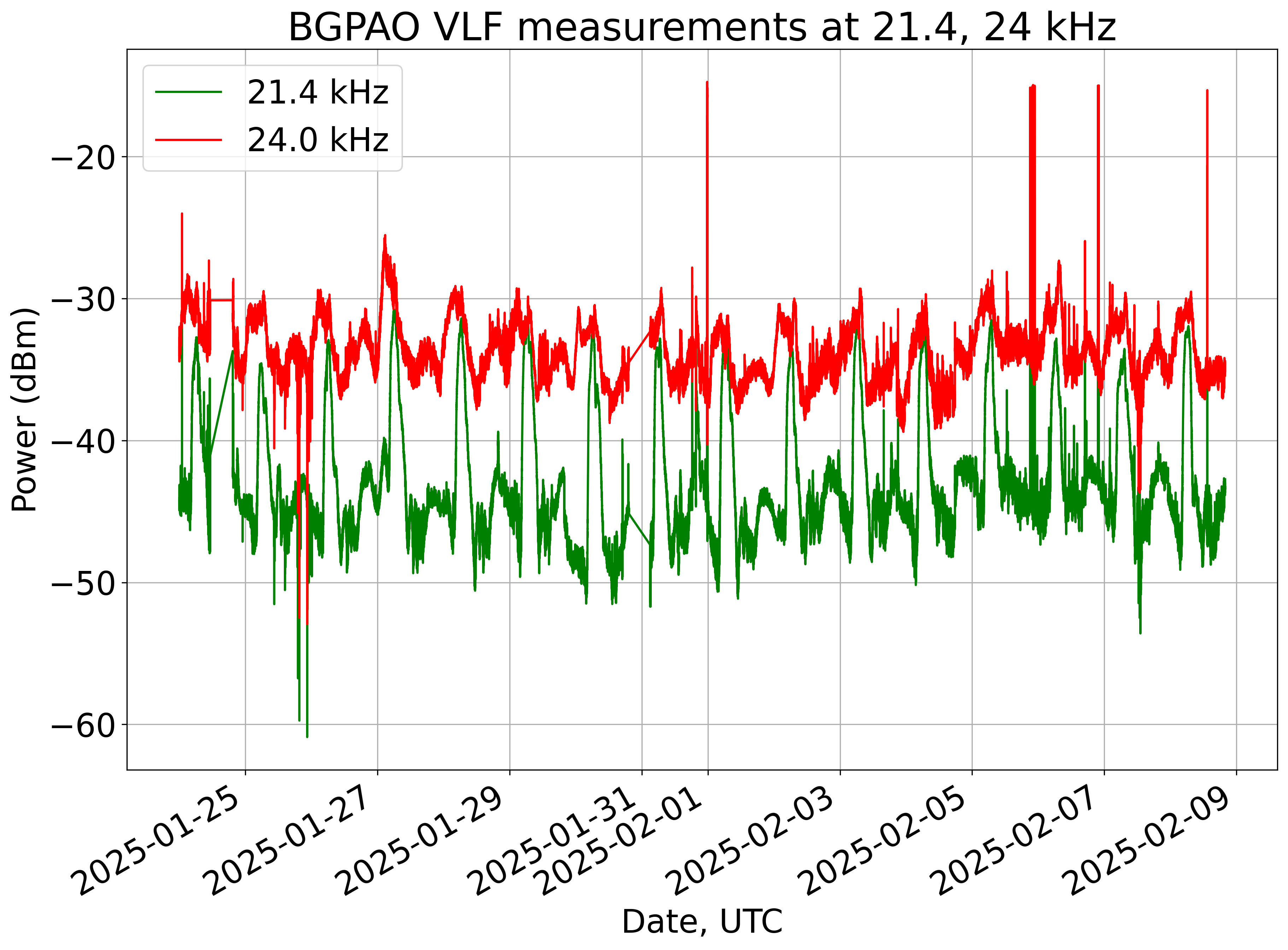}
\caption{The VLF observations obtained from the St. Kliment Ohridski Base during the period 01/24/2025 -- 02/08/2025 at 21.4 kHz (green) and 24.0 kHz (red).}
\label{fig_vlf_overview}
\end{figure*}

We analyze the VLF observations performed in the period 01/24/2025 -- 02/08/2025, which are shown in Figure \ref{fig_vlf_overview}. The daily variations in the observed signal are readily visible. Signal was not observed late on 01/25/2025 and late on 01/31/2025--early 02/01/2025 due to power outages. We studied the relationship between X-ray and VLF observations of 250 flares of class C and M during the period (no X-class flares occurred during it). Table \ref{flare_list} provides information about the flares from the NOAA/SWPC catalog of GOES-observed flares.

The present analysis focuses on empirical detectability rather than detailed waveguide inversion or propagation modeling. Consequently, the observed VLF perturbations should be interpreted as path-integrated responses whose amplitude and timing may depend on modal structure, propagation geometry, and ground conductivity variations along the transmitter-receiver path. In future work, we will address these effects.

\begin{figure*}[t]
\centering
\includegraphics[width=0.8\textwidth]{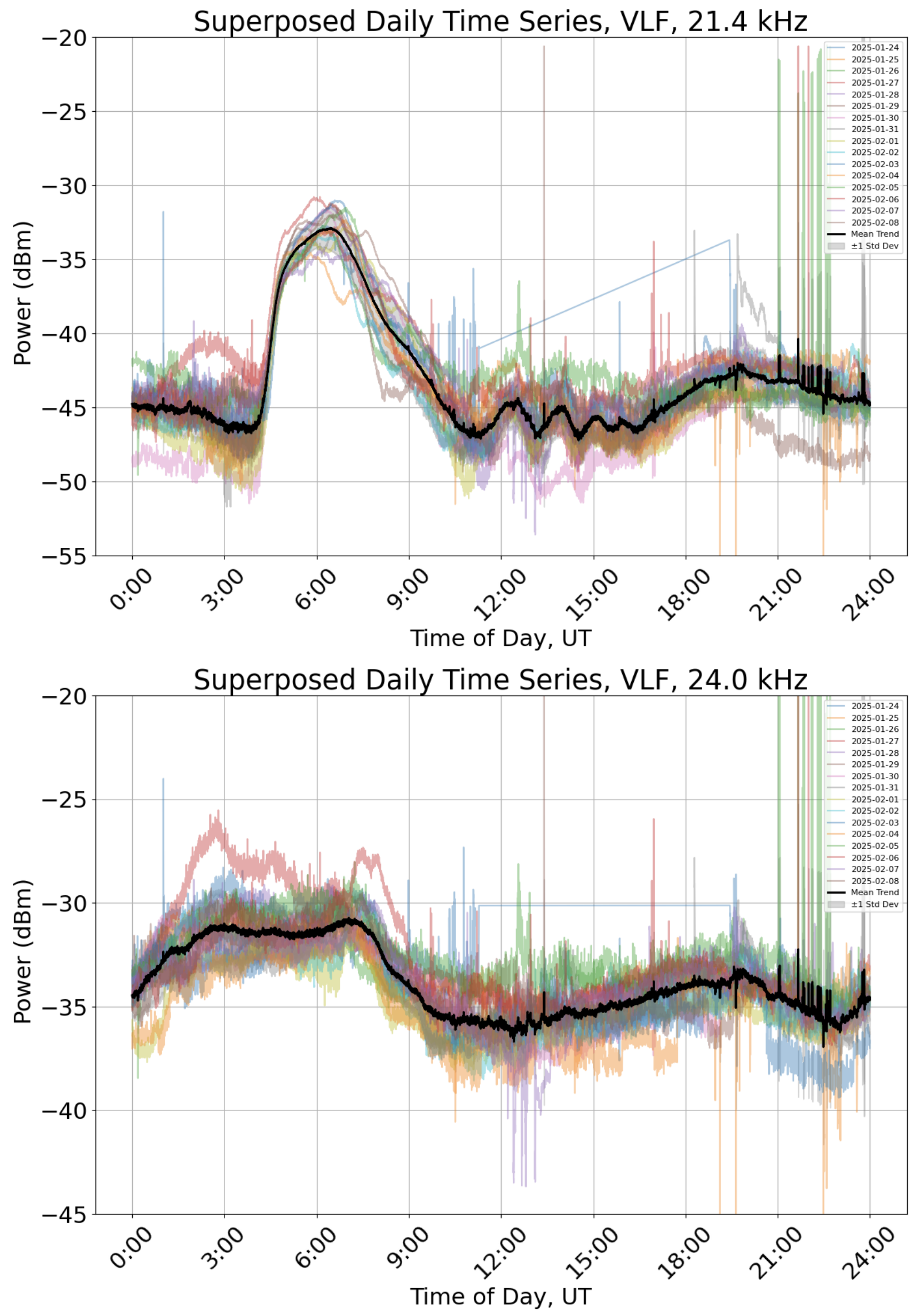}
\caption{Daily superposed time series of 21.4 kHz (top) and 24.0 kHz (bottom) VLF observations. Different colors are used for each day. The black lines show the mean for each time step, and the gray shadings denote the standard deviations about the mean.}
\label{fig_superposed_epoch_analysis}
\end{figure*}

In order to detect changes to the observed signal due to solar activity, we first detrended the signal via superposed epoch analysis. Figure \ref{fig_superposed_epoch_analysis} shows the superposed daily signals in the two frequencies. The daily trend is clearly visible in both, peaking shortly after 6h UT. The means of the superposed daily time series are shown in black, and the gray shadings show the standard deviations about it. The diurnal trend is more pronounced in the lower frequencies.

The strong diurnal variation in the VLF signal was removed using a superposed-epoch approach applied separately to the 21.4 and 24.0~kHz time series. For each frequency, the observations were grouped by time of day, and a mean daily profile was constructed from the available measurements at each time step. Days containing incomplete observing intervals were excluded from the construction of the mean reference profile, but were retained in the final detrended dataset. Data gaps caused by power outages or other missing intervals were ignored when computing the mean profile, and no interpolation or gap filling was applied. Consequently, each point of the mean daily profile was calculated only from the measurements actually available at that time of day. The resulting frequency-specific mean profile was then subtracted from every corresponding daily time series, including days that had been excluded from the averaging step. This procedure removes the repeatable diurnal component while preserving shorter-timescale variations, including possible flare-related perturbations.

\begin{figure*}[t]
\centering
\includegraphics[width=0.8\textwidth]{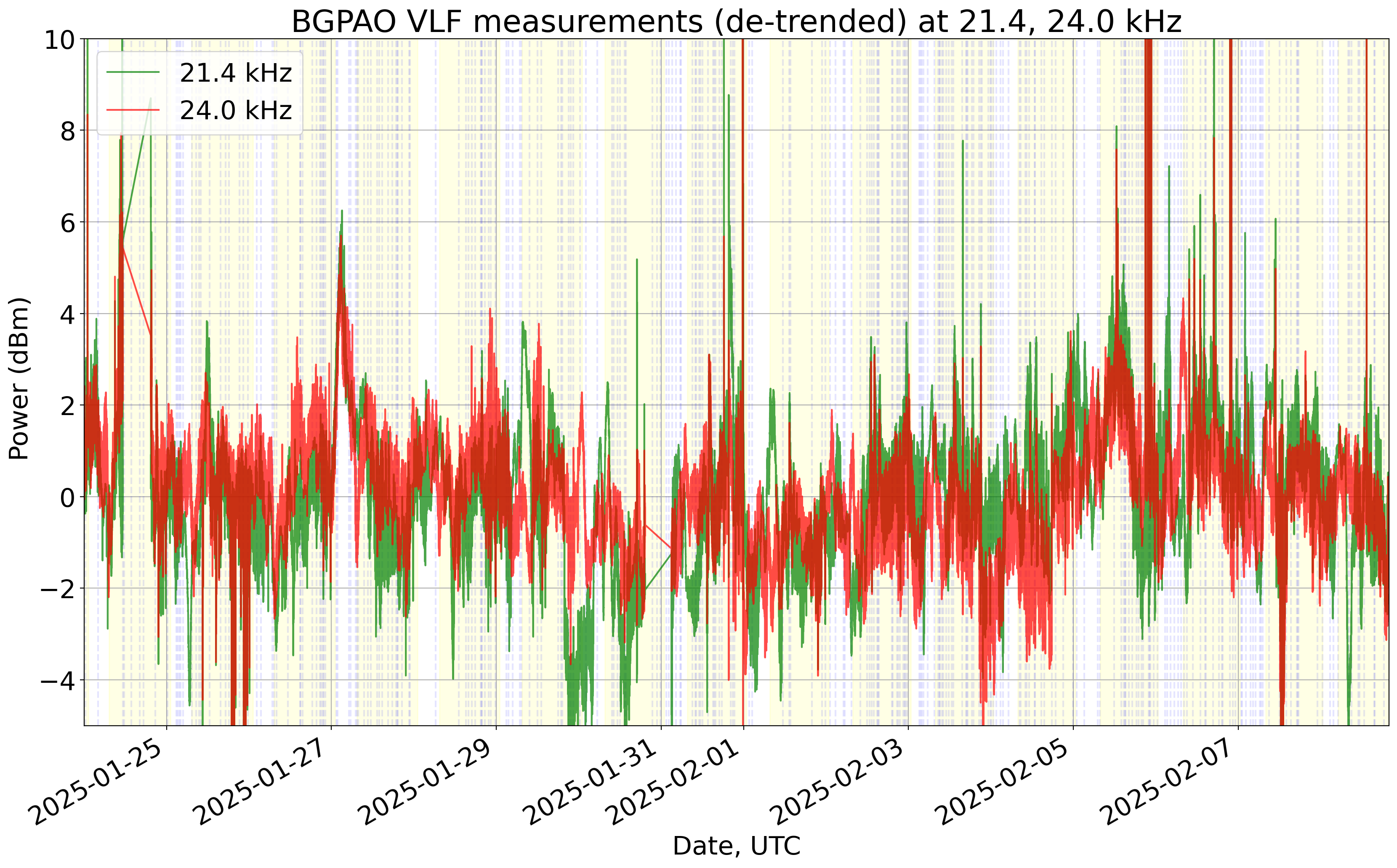}
\caption{Detrended VLF signal in 21.4 kHz (green) and 24.0 kHz (red) during the studied period. The local daytime periods are shaded in yellow. Vertical dashed lines denote the starting times of $>$C7.0 flares.}
\label{fig_detrended_daytimes}
\end{figure*}

We next subtracted the daily trends from the two signals, in order to seek weaker signals related to solar flares. Figure \ref{fig_detrended_daytimes} shows the de-trended signals for the two frequencies with superposed the daytime periods in yellow and the start times of the relatively strong flares ($>$C7.0) as vertical dashed lines.

\begin{figure*}[t]
\centering
\includegraphics[width=0.8\textwidth]{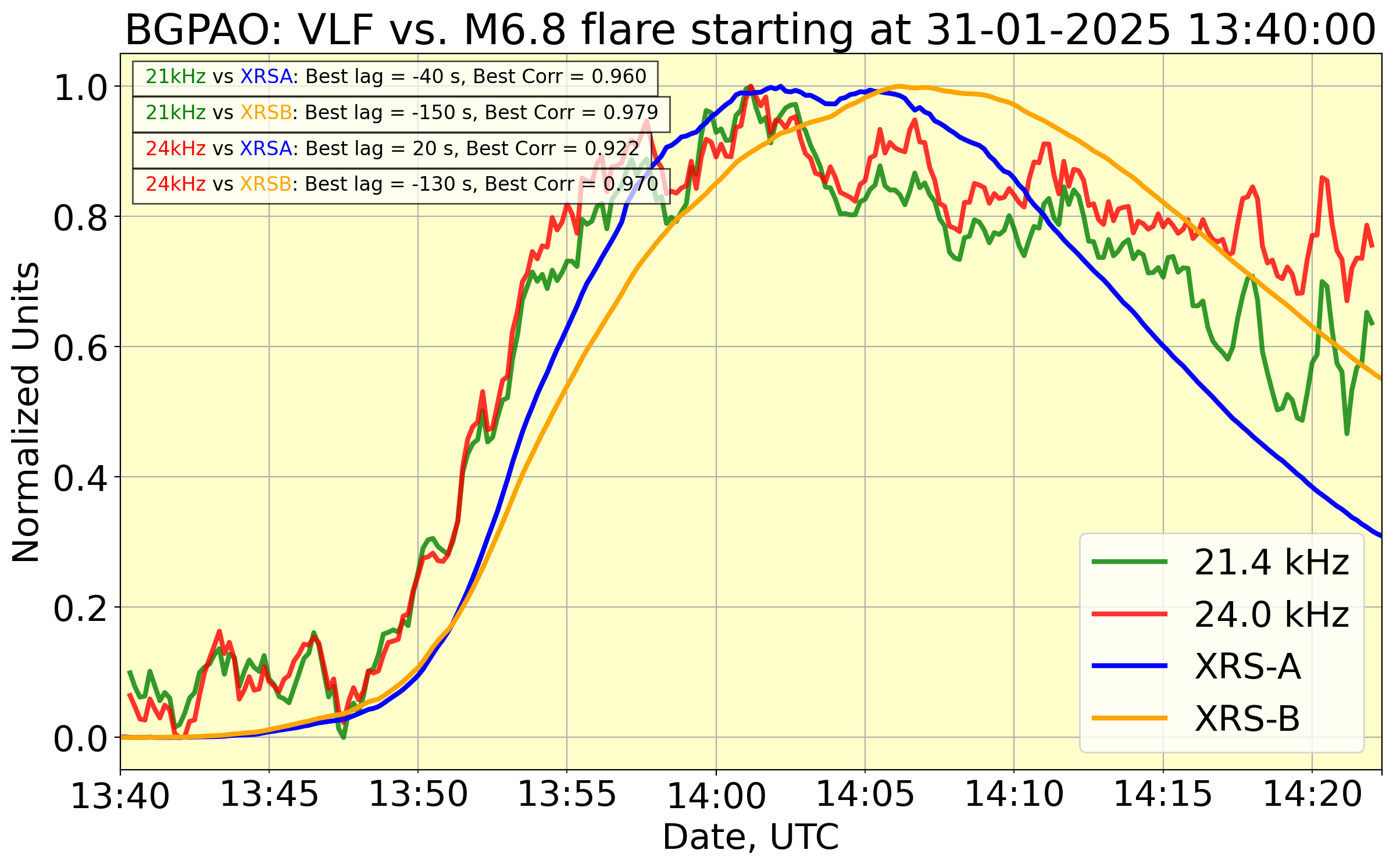}
\caption{Comparison between the GOES X-ray and BGPAO VLF observations during a flare on 31.01.2025.}
\label{fig_flare_31012025}
\end{figure*}

We distinguish between three different timing quantities used in this work: (i) the physical ionospheric response delay, which represents the causal delay between enhanced solar X-ray flux and the resulting D-region modification; (ii) the correlation lag $\Delta t$ obtained from cross-correlation analysis, which quantifies the temporal shift yielding maximum profile similarity between the VLF and X-ray time series; and (iii) the peak-to-extremum delay $\Delta t_{\rm peak}$, defined as the difference between the times of the X-ray peak and the corresponding VLF response extremum. Only the latter two quantities are directly measured in this study. Negative or near-zero correlation lags should therefore not be interpreted directly as negative physical ionospheric response times.

\subsection{Correlation Analysis}
\label{Correlation Analysis}

We perform a cross-correlation analysis between each VLF signal and the two GOES X-ray channels. Because flare-induced VLF amplitude perturbations can have either sign,
we treat positive and negative VLF--X-ray correlations equivalently. For the correlation-lag characterization, we evaluate the full signed cross-correlation function over a $\pm360$~s lag interval for each VLF--X-ray pairing. We retain both the maximum and minimum values of the cross-correlation function,
\begin{equation}
\rho_{\max} = \max_{\tau}\rho(\tau),
\qquad
\rho_{\min} = \min_{\tau}\rho(\tau),
\end{equation}
and define the dominant signed correlation extremum, $\rho_*$, as
\begin{equation}
\rho_* =
\begin{cases}
\rho_{\max}, & |\rho_{\max}| \geq |\rho_{\min}|,\\
\rho_{\min}, & |\rho_{\min}| > |\rho_{\max}|.
\end{cases}
\end{equation}
Thus, $\rho_*$ retains the sign of the correlation extremum with the largest absolute magnitude. The corresponding lag is denoted $\Delta t_*$. For the operational detection analysis, the same quantities are evaluated within the $\pm180$~s association interval, and a VLF--X-ray pairing is classified as a detection when $|\rho_*|\geq0.7$. Thus, positive- and negative-polarity flare-related VLF responses enter the detection criterion symmetrically.

Separately from the operational detection calculation, the signed cross-correlation analysis is evaluated over the wider $\pm360$~s lag interval in order to characterize the correlation-lag behaviour. Correlation extrema occurring at the limits of the search interval, $|\Delta t_*|=360$~s, are classified as boundary-limited because the true extremum may lie outside the investigated range; these cases are retained in Table~\ref{tab_correlation} for completeness but are not interpreted as resolved lag measurements.

The VLF response time is defined consistently with the sign of $\rho_{*}$. For positively correlated responses ($\rho_{*}>0$), the relevant VLF response extremum is the maximum detrended VLF amplitude within the flare interval, whereas for negatively correlated responses ($\rho_{*}<0$), it is the minimum detrended amplitude. We therefore refer to this quantity as the VLF extremum rather than universally as the VLF peak. The corresponding VLF--X-ray delay statistics are calculated using the appropriate signed extremum for each detected pairing.

\subsection{Path-integrated illumination fraction along the propagation path}
\label{path_illumination}

Because VLF propagation is sensitive to the Earth--ionosphere waveguide integrated over the full transmitter--receiver path, receiver-local day/night classification is only a coarse proxy for the conditions controlling flare detectability. We therefore quantify illumination geometry using the fraction of the propagation path that is sunlit at the flare time. The transmitter--receiver path is represented by a sequence of latitude--longitude points sampled along the great-circle trajectory. At each point $i$, the solar zenith angle $\chi_i(t)$ is computed from standard solar-position geometry at time $t$. A point is classified as sunlit if $\chi_i(t) < 90^\circ$ (equivalently $\cos\chi_i(t) > 0$).

We define the sunlit path fraction as the path-length-weighted mean of this binary classification,
\begin{equation}
f_{\rm sun}(t) \approx \frac{\sum_{j=1}^{N-1} H\!\left(\cos\chi_{j}(t)\right)\, \Delta s_j}{\sum_{j=1}^{N-1} \Delta s_j},
\end{equation}
where $H$ is the Heaviside step function, $\Delta s_j$ is the great-circle distance of segment $j$ between consecutive path points, and the illumination state of each segment is taken from the adjacent sampled points (equivalently from a segment midpoint). By construction, $f_{\rm sun}=1$ corresponds to a fully sunlit path and $f_{\rm sun}=0$ to a fully dark path. This metric provides a simple first-order characterization of illumination geometry along the propagation path. The binary illumination formalism does not capture the continuous dependence of D-region ionization on solar zenith angle and should therefore be interpreted as a qualitative indicator rather than a rigorous physical parameterization. For the binned analysis presented below, we classify a propagation path as dark for $0\leq f_{\rm sun}\leq0.2$, mixed for $0.2<f_{\rm sun}\leq0.8$, and sunlit for $0.8<f_{\rm sun}\leq1$.

\subsection{Operational definition of VLF flare detection and detection efficiency}

To assess the feasibility of using long-baseline VLF observations as a practical detector of solar flares, we adopt an operational definition of flare detectability based on the temporal correspondence between the de-trended VLF signal and the GOES soft X-ray flux. For each flare and each VLF--X-ray pairing (21.4~kHz or 24.0~kHz with either GOES XRS channel), we evaluate the cross-correlation function within
a symmetric operational association window of $\pm180$~s. Within this interval, we determine the maximum and minimum correlation coefficients,
\begin{equation}
\rho_{\max} =
\max_{|\tau|\leq180\,{\rm s}} \rho(\tau),
\end{equation}
and
\begin{equation}
\rho_{\min} =
\min_{|\tau|\leq180\,{\rm s}} \rho(\tau).
\end{equation}
The dominant signed correlation extremum, $\rho_{*}$, is defined as the coefficient with the larger absolute magnitude,
\begin{equation}
\rho_{*} =
\begin{cases}
\rho_{\max}, & |\rho_{\max}| \geq |\rho_{\min}|,\\
\rho_{\min}, & |\rho_{\min}| > |\rho_{\max}|.
\end{cases}
\end{equation}
The sign of $\rho_{*}$ therefore retains the polarity of the dominant
VLF--X-ray response.

A VLF--X-ray pairing is classified as a detection when
\begin{equation}
|\rho_*| \geq \rho_0,
\end{equation}
where we adopt $\rho_0=0.7$. Thus, positive- and negative-polarity VLF responses enter the operational detection criterion on an equal basis.

A flare is considered detected at a given VLF frequency if the criterion is satisfied for at least one of the two GOES XRS channels. It is considered detected overall if the criterion is satisfied for at least one of the four VLF--X-ray pairings. Detection at both VLF frequencies requires that at least one XRS pairing satisfy the criterion independently at 21.4~kHz and at 24.0~kHz. The $\pm180$~s interval is incorporated directly into the operational definition and is therefore not applied as an additional lag criterion. It represents the adopted temporal association window for linking a VLF perturbation with the corresponding GOES flare. This operational detection calculation is distinct from the wider $\pm360$~s signed cross-correlation analysis described in Sect.~\ref{Correlation Analysis}, which is used to characterize correlation lags and identify boundary-limited extrema outside the operational detection interval.

Using this definition, we compute the VLF flare detection efficiency as
\begin{equation}
\eta = \frac{N_{\mathrm{det}}}{N_{\mathrm{total}}},
\end{equation}
where $N_{\mathrm{det}}$ is the number of flares satisfying the relevant detection criterion and $N_{\mathrm{total}}$ is the total number of flares in the corresponding sample. Detection efficiencies are evaluated for the full sample and separately by flare class, receiver-local day/night classification, propagation-path illumination, VLF frequency, and individual VLF--X-ray pairing. We also report the more restrictive case in which a flare is detected at both VLF frequencies. This formulation provides a quantitative measure of the performance and limitations of long-baseline VLF observations as a proxy for solar flare occurrence.

%\subsection{Flare Detection}
%VLF onset times were determined using an algorithm anchored to the GOES soft X-ray flare onset, whereby baseline estimation and onset detection were constrained to a physically plausible temporal window around the X-ray start time. The X-ray peak time was defined as the time of maximum X-ray flux within the GOES-defined flare interval. The corresponding VLF response time was defined according to the polarity of the detected response.

\section{Results}
\label{results}

Table~\ref{tab_detections} summarizes the detection efficiencies for the full flare sample and separately for C- and M-class events, reported at three levels: overall detection in any VLF--X-ray pairing, detection at each VLF frequency regardless of XRS channel, and detection in the four individual VLF--X-ray pairings. Across the full sample of 250 flares, 203 events are detected in at least one pairing, corresponding to an overall detection efficiency of 81.2\%. At the frequency level, 177/250 events (70.8\%) are detected at 21.4~kHz and 155/250 (62.0\%) at 24.0~kHz. Detection at both VLF frequencies occurs for 129/250 events (51.6\%), while 74/250 (29.6\%) are detected at only one of the two frequencies.

Among the 203 flares detected in at least one pairing, 177 (87.2\%) are detected at 21.4~kHz, 155 (76.4\%) at 24.0~kHz, and 129 (63.5\%) at both frequencies. Thus, the 21.4~kHz channel has the higher observed detection efficiency in the present dataset, while the substantial overlap between the two frequencies shows that many flare responses are detected independently along both propagation paths. Because the two frequencies correspond to different transmitter--propagation-path--receiver channels, the difference in their detection efficiencies should not be interpreted as an isolated frequency effect. At the level of individual VLF--X-ray pairings, the highest detection efficiency is obtained for 21.4~kHz--XRS-B, with 165/250 detections (66.0\%), followed by 21.4~kHz--XRS-A with 151/250 (60.4\%), 24.0~kHz--XRS-B with 146/250 (58.4\%), and 24.0~kHz--XRS-A with 134/250 (53.6\%). The class-dependent differences underlying these aggregate values are examined in Sect.~3.1.

The signed-correlation analysis further shows that negative-polarity responses constitute a substantial fraction of the observed VLF--X-ray behaviour. Across the four pairings and 250 flares, 428 of the 1000 full-range correlation functions (42.8\%) have a negative dominant correlation extremum, compared with 572 (57.2\%) with a positive extremum. The fractions of negative dominant extrema are 93/250 (37.2\%) for 21.4~kHz--XRS-A, 102/250 (40.8\%) for 21.4~kHz--XRS-B, 121/250 (48.4\%) for 24.0~kHz--XRS-A, and 112/250 (44.8\%) for 24.0~kHz--XRS-B. Among the 751 pairings with $|\rho_*|\geq0.7$ in the extended $\pm360$~s analysis, 301 (40.1\%) have negative polarity. Of the 1000 full-range correlation functions, 234 (23.4\%) have their dominant signed extremum at one of the $\pm360$~s search boundaries. These boundary-limited cases are retained in Table~\ref{tab_correlation} for completeness but are not interpreted as resolved correlation-lag measurements.

%\begin{table}
%\centering
%\label{tab_detections}
%\begin{tabular}{lccccccc}
%\toprule
%Sample & Any & 21 kHz & 24 kHz & 21-A & 21-B & 24-A & 24-B \\
%\midrule
%All flares & 60.8\% & 52.4\% & 40.4\% & 41.6\% & 47.6\% & 33.2\% & 36.8\% \\
%C-class    & 56.6\% & 46.7\% & 36.3\% & 35.8\% & 41.0\% & 28.8\% & 32.5\% \\
%M-class    & 84.2\% & 84.2\% & 63.2\% & 73.7\% & 84.2\% & 57.9\% & 60.5\% \\
%\bottomrule
%\end{tabular}
%\caption{Detection efficiencies (\%) by frequency and radio--X-ray pairing for the full sample and by flare class. ``Any'' denotes detection in any frequency and any XRS channel. ``21 kHz (any)'' and ``24 kHz (any)'' denote detection at the respective frequency regardless of XRS channel.}
%\end{table}

\begin{table*}
\centering
\begin{tabular}{lccccccc}
\toprule
Sample & Any & 21.4 kHz (any) & 24.0 kHz (any) & 21A & 21B & 24A & 24B \\
\midrule
All (250) 
& 81.2 (203) 
& 70.8 (177) 
& 62.0 (155) 
& 60.4 (151) 
& 66.0 (165) 
& 53.6 (134) 
& 58.4 (146) \\

C-class (212) 
& 78.8 (167) 
& 67.0 (142) 
& 59.4 (126) 
& 56.1 (119) 
& 61.3 (130) 
& 50.5 (107) 
& 55.7 (118) \\

M-class (38) 
& 94.7 (36) 
& 92.1 (35) 
& 76.3 (29) 
& 84.2 (32) 
& 92.1 (35) 
& 71.1 (27) 
& 73.7 (28) \\
\bottomrule
\end{tabular}

\caption{Detection efficiencies (\%) by frequency and VLF--X-ray pairing for the full sample and by flare class, using the signed operational detection criterion $|\rho_*|\geq0.7$ within the $\pm180$~s association window. ``Any'' denotes detection in at least one VLF--X-ray pairing. ``21.4 kHz (any)'' and ``24.0 kHz (any)'' denote detection at the respective VLF frequency in at least one GOES XRS channel. The number of detected events is shown in parentheses.}

\label{tab_detections}
\end{table*}

\begin{table*}
\centering
\begin{tabular}{lllrcccc}
\toprule
Local time & $f_{\rm sun,21}$ & $f_{\rm sun,24}$ & $N$
& $\eta_{\rm any}$ & $\eta_{21}$ & $\eta_{24}$ & $\eta_{\rm both}$ \\
\midrule

Night & dark   & dark   & 27 & 96.3 & 96.3 & 55.6 & 55.6 \\
Night & mixed  & dark   & 15 & 80.0 & 53.3 & 53.3 & 26.7 \\
Night & sunlit & dark   & 13 & 100.0 & 76.9 & 92.3 & 69.2 \\

Day   & dark   & dark   & 12 & 83.3 & 66.7 & 66.7 & 50.0 \\
Day   & dark   & mixed  & 16 & 50.0 & 37.5 & 37.5 & 25.0 \\
Day   & dark   & sunlit & 1  & 100.0 & 100.0 & 100.0 & 100.0 \\
Day   & mixed  & sunlit & 49 & 79.6 & 71.4 & 61.2 & 53.1 \\
Day   & sunlit & dark   & 5  & 100.0 & 60.0 & 80.0 & 40.0 \\
Day   & sunlit & mixed  & 13 & 69.2 & 61.5 & 61.5 & 53.8 \\
Day   & sunlit & sunlit & 61 & 72.1 & 60.7 & 55.7 & 44.3 \\

\bottomrule
\end{tabular}
\caption{Detection efficiencies (\%) for C-class flares as a function of receiver-local day/night classification and path-integrated illumination. The illumination
regimes are defined from the sunlit path fraction as dark ($0\leq f_{\rm sun}\leq0.2$), mixed ($0.2<f_{\rm sun}\leq0.8$), and sunlit ($0.8<f_{\rm sun}\leq1$). $\eta_{\rm any}$ denotes detection in at least one VLF--X-ray pairing, $\eta_{21}$ and $\eta_{24}$ denote detection at 21.4 and 24.0~kHz, respectively, in at least one XRS channel, and $\eta_{\rm both}$ denotes detection at both VLF frequencies. Detection is based on the signed operational criterion $|\rho_*|\geq0.7$ within the $\pm180$~s association window.}
\label{tab_detection_c}
\label{flare_detections_c}
\end{table*}

\begin{table*}
\centering
\begin{tabular}{lllrcccc}
\toprule
Local time & $f_{\rm sun,21}$ & $f_{\rm sun,24}$ & $N$
& $\eta_{\rm any}$ & $\eta_{21}$ & $\eta_{24}$ & $\eta_{\rm both}$ \\
\midrule

Night & dark   & dark   & 9 & 100.0 & 100.0 & 66.7 & 66.7 \\
Night & mixed  & dark   & 4 & 50.0  & 50.0  & 25.0 & 25.0 \\
Night & sunlit & dark   & 2 & 100.0 & 50.0  & 100.0 & 50.0 \\
Day   & dark   & dark   & 4 & 100.0 & 100.0 & 50.0 & 50.0 \\
Day   & dark   & mixed  & 2 & 100.0 & 100.0 & 100.0 & 100.0 \\
Day   & dark   & sunlit & 1 & 100.0 & 100.0 & 100.0 & 100.0 \\
Day   & mixed  & sunlit & 8 & 100.0 & 100.0 & 100.0 & 100.0 \\
Day   & sunlit & mixed  & 3 & 100.0 & 100.0 & 100.0 & 100.0 \\
Day   & sunlit & sunlit & 5 & 100.0 & 100.0 & 80.0  & 80.0 \\

\bottomrule
\end{tabular}
\caption{Detection efficiencies (\%) for M-class flares as a function of receiver-local day/night classification and path-integrated illumination.}
\label{flare_detections_m}
\end{table*}

\subsection{Detection Efficiency by Flare Class}

Tables~\ref{flare_detections_c} and~\ref{flare_detections_m} summarize the detection efficiencies for C- and M-class flares, respectively, after applying
the signed operational detection criterion described in Sect.~\ref{methods}. In the full sample, M-class flares show a higher overall detection efficiency
than C-class flares: 36/38 (94.7\%) of the M-class events are detected in at least one VLF--X-ray pairing, compared with 167/212 (78.8\%) of the C-class
events. The same tendency is present at both VLF frequencies. At 21.4~kHz, 35/38 (92.1\%) of the M-class events and 142/212 (67.0\%) of the C-class
events are detected, while at 24.0~kHz the corresponding fractions are 29/38 (76.3\%) and 126/212 (59.4\%), respectively. Dual-frequency detections
are also more frequent for the M-class sample, occurring for 28/38 (73.7\%) events compared with 101/212 (47.6\%) for C-class flares.

The illumination-stratified results in Tables~\ref{flare_detections_c} and~\ref{flare_detections_m} are broadly consistent with this overall
difference, although several individual bins contain only a small number of events. For example, under nighttime dark--dark conditions, all 9 M-class
events are detected in at least one pairing, compared with 26/27 (96.3\%) C-class events. Under nighttime mixed--dark conditions, however, the M-class
efficiency is 2/4 (50.0\%), whereas the C-class efficiency is 12/15 (80.0\%). During local daytime, all of the populated M-class illumination bins have
100\% ``any-pairing'' detection, but several of these categories contain only one to five events. These percentages should therefore be interpreted as
descriptive properties of the present sample rather than as statistically established differences between individual illumination regimes. The statistics nevertheless show a clear sample-level association between flare magnitude and detectability: stronger M-class events are more frequently detected and more often confirmed at both VLF frequencies than C-class events. A larger sample will be required to determine how this flare-class dependence interacts quantitatively with propagation-path illumination and other observing conditions.

\subsection{Detection Efficiency by Time of Day}

The receiver-local day/night classification provides a simple description of the observing geometry, but the revised statistics do not indicate a single
monotonic day--night dependence of VLF detectability. Instead, the observed efficiencies vary with both flare class and the illumination state of the two
propagation paths.

For C-class flares, nighttime ``any-pairing'' efficiencies range from 80.0\% (12/15) for mixed--dark paths to 100.0\% (13/13) for sunlit--dark paths,
with 96.3\% (26/27) obtained for dark--dark conditions. During local daytime, the corresponding efficiencies span a wider range, from 50.0\% (8/16) for
dark--mixed conditions to 100\% in the dark--sunlit and sunlit--dark bins, although the latter contain only 1 and 5 events, respectively. The two most
populated daytime categories yield 79.6\% (39/49) for mixed--sunlit paths and 72.1\% (44/61) for sunlit--sunlit paths.

For M-class flares, the nighttime efficiencies are 100\% (9/9) for dark--dark, 50.0\% (2/4) for mixed--dark, and 100\% (2/2) for
sunlit--dark conditions. All populated daytime M-class bins have 100\% ``any-pairing'' detection. Dual-frequency confirmation during daytime
is also frequent, ranging from 50.0\% (2/4) in the dark--dark bin to 100\% in several illumination categories. However, the M-class subsamples are
small, particularly after subdivision by both local time and path illumination.

Thus, receiver-local time of day alone does not provide a sufficient description of the detection efficiency. The observed variations are more
appropriately interpreted together with the illumination state of the full transmitter--receiver paths. Given the limited number of events in several
bins, we do not assign statistical significance to the differences between individual day and night categories.

\subsection{Detection Efficiency by Path Illumination}

The path-integrated sunlit fraction provides additional information on the propagation geometry beyond the receiver-local day/night classification.
Tables~\ref{flare_detections_c} and~\ref{flare_detections_m} show the detection efficiencies after classifying each propagation path as dark,
mixed, or sunlit according to the $f_{\rm sun}$ intervals defined in Sect.~\ref{path_illumination}. The resulting binned statistics show
substantial differences among illumination configurations, but they do not exhibit a simple monotonic progression from dark to sunlit conditions.

For C-class flares observed at night, the ``any-pairing'' detection efficiency is 96.3\% (26/27) when both paths are dark, 80.0\% (12/15) when
the 21.4~kHz path is mixed and the 24.0~kHz path is dark, and 100.0\% (13/13) when the 21.4~kHz path is sunlit while the 24.0~kHz path remains
dark. The corresponding dual-frequency efficiencies are 55.6\%, 26.7\%, and 69.2\%, respectively. During local daytime, the two largest C-class
subsamples yield 79.6\% (39/49) ``any-pairing'' detection for mixed--sunlit conditions and 72.1\% (44/61) for sunlit--sunlit conditions, with
dual-frequency detection rates of 53.1\% and 44.3\%, respectively.

For M-class flares, high detection efficiencies are observed in several illumination regimes, but the number of events in the individual bins is
small. The dark--dark nighttime sample contains 9 events and has 100\% ``any-pairing'' detection and 66.7\% dual-frequency detection, whereas the
mixed--dark nighttime sample contains only 4 events and yields corresponding efficiencies of 50.0\% and 25.0\%. During daytime, all populated M-class bins
reach 100\% ``any-pairing'' detection, while dual-frequency confirmation ranges from 50.0\% to 100\%.

These results show that path illumination is useful for organizing the observed variation in VLF detectability, but the present sample does not
support a statistically robust inference that a particular illumination state systematically enhances or suppresses detection. In particular, the
small populations of several bins make individual percentages highly sensitive to one or two events. We therefore treat the illumination results
as descriptive sample patterns. A larger event set, preferably analysed using the continuous $f_{\rm sun}$ values together with flare magnitude,
frequency, and local time, will be required to quantify the independent contribution of illumination geometry to flare detectability.

\subsection{Statistical estimation of the $\Delta t_{peak}$ parameter}

\begin{figure*}[t]
\centering
\includegraphics[width=1.\textwidth]{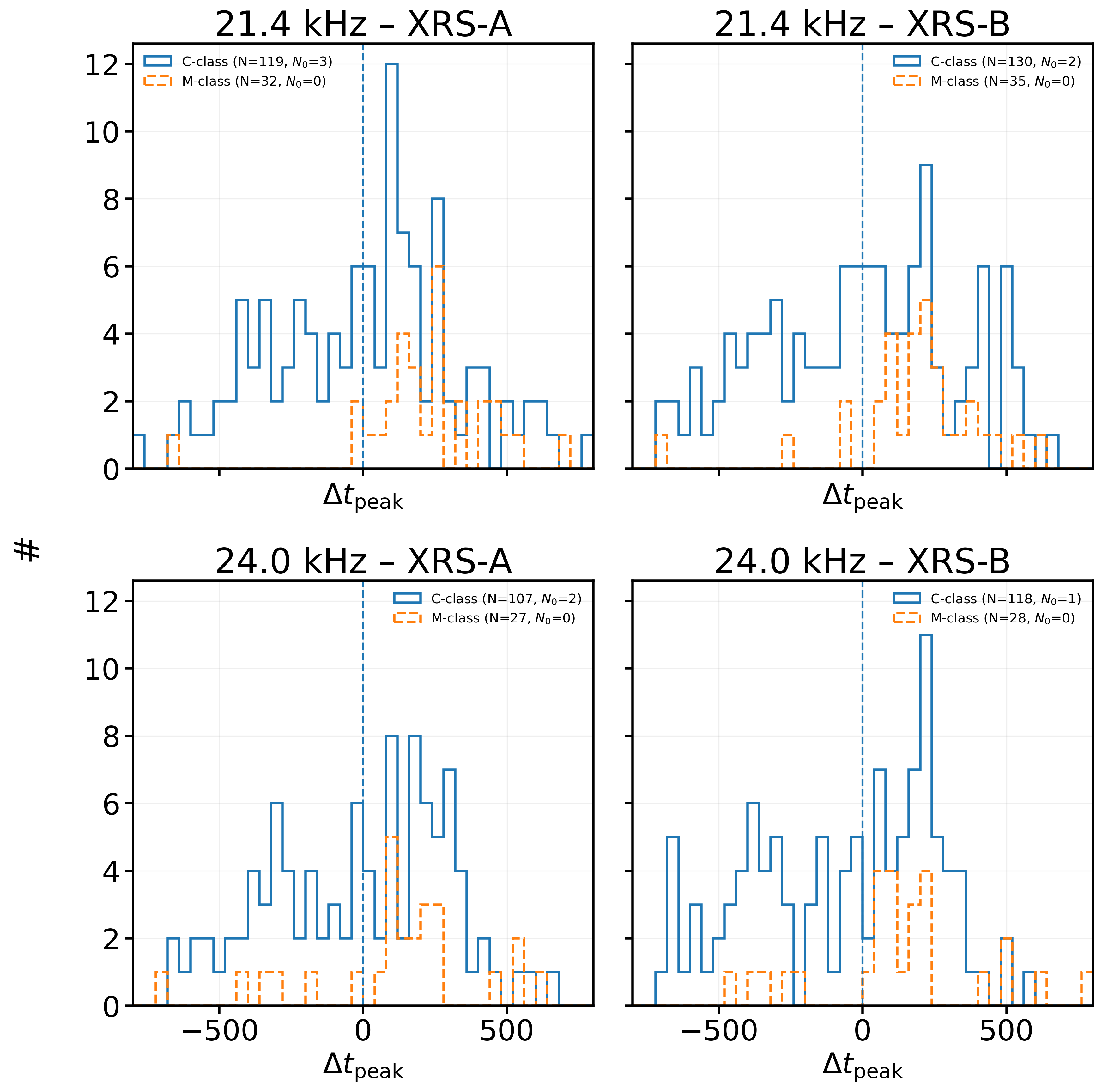}
\caption{Histograms of the distributions of the $\Delta t_{\rm peak}=t_{\rm X-ray}-t_{\rm VLF}$ parameter for detected C-class and M-class flares. Only events satisfying the detection criterion for the corresponding VLF--X-ray pairing are included. Negative values indicate that the X-ray peak occurs first, while positive values indicate that the VLF response extremum occurs first. $N_0$ denotes exact zero-delay cases. Because individual flares may contribute to more than one panel, the four pairing samples are not statistically independent.}
\label{fig_dt_distributions}
\end{figure*}

In addition to the correlation between the full X-ray and VLF time series of each flare, we examined the distributions of the time differences between the X-ray peak and the corresponding VLF response extremum (without shifting either time series), denoted as $\Delta t_{\rm peak}$. The VLF response extremum is defined according to the polarity of the detected response: the maximum detrended VLF amplitude is used for a positive correlation and the minimum detrended amplitude for a negative correlation. The delay is defined as $\Delta t_{\rm peak}=t_{\rm X-ray}-t_{\rm VLF}$, so that negative values indicate that the X-ray peak occurs first and the VLF response follows, whereas positive values indicate that the VLF extremum precedes the X-ray peak. The statistics are calculated separately for each VLF--X-ray pairing and only for events satisfying the operational detection criterion for that specific pairing, as defined in Sect.~\ref{methods}. Consequently, the number of delay measurements is identical to the corresponding number of detections: for C-class flares the 21A, 21B, 24A, and 24B samples contain 119, 130, 107, and 118 events, respectively, while for M-class flares they contain 32, 35, 27, and 28 events. Zero-delay values are retained in the analysis rather than assigned to either sign. The pairing-specific distributions shown in Fig.~\ref{fig_dt_distributions} are summarised statistically in Table~\ref{tab_dt_stats}.

The revised delay statistics show a different behaviour for C- and M-class flares. For C-class events, the mean $\Delta t_{\rm peak}$ values are relatively close to zero and differ in sign among the four pairings, ranging from approximately $-51$ to $+45$~s. The corresponding numbers of VLF-earlier and VLF-later events are also broadly comparable, indicating substantial variability in the relative timing of the two responses. In contrast, all four M-class pairings have positive mean delays, ranging from approximately $+135$ to $+194$~s, and VLF-earlier cases clearly outnumber VLF-later cases: 27 versus 3 for 21A, 29 versus 6 for 21B, 21 versus 6 for 24A, and 23 versus 5 for 24B. The large standard deviations, of order several hundred seconds, nevertheless demonstrate considerable event-to-event variability in both flare classes. These results therefore do not imply a single universal ordering of the VLF and soft X-ray extrema, although the M-class sample shows a stronger tendency for the VLF extremum to precede the X-ray maximum. Because an individual flare can satisfy the detection criterion for more than one VLF--X-ray pairing, the four delay distributions contain repeated measurements of some flares and are not statistically independent. They are therefore presented and interpreted as separate pairing-specific distributions rather than combined into a single sample. A more detailed investigation of the physical origin of the delay distributions, including additional X-ray spectral information and possible relationships between spectral hardness and $\Delta t_{\rm peak}$, is left for future work.

\begin{table*}
\centering
\begin{tabular}{llrrrrrrr}
\toprule
Flare class & Pair &
$N$ &
$\langle \Delta t_{\rm peak} \rangle$ [s] &
$\sigma(\Delta t_{\rm peak})$ [s] &
$N_{\rm VLF\,later}$ &
$N_{\rm co}$ &
$N_{\rm VLF\,earlier}$ &
$N_{0}$ \\
\midrule

C-class & 21A & 119 &   6 & 409 & 49 & 10 & 60 & 3 \\
C-class & 21B & 130 & -51 & 462 & 63 &  6 & 61 & 2 \\
C-class & 24A & 107 &  45 & 405 & 44 &  7 & 56 & 2 \\
C-class & 24B & 118 & -36 & 433 & 57 &  2 & 59 & 1 \\

\midrule

M-class & 21A & 32 & 194 & 374 & 3 & 2 & 27 & 0 \\
M-class & 21B & 35 & 138 & 439 & 6 & 0 & 29 & 0 \\
M-class & 24A & 27 & 135 & 337 & 6 & 0 & 21 & 0 \\
M-class & 24B & 28 & 146 & 319 & 5 & 0 & 23 & 0 \\

\bottomrule
\end{tabular}

\caption{Summary statistics of the pairing-specific $\Delta t_{\rm peak}=t_{\rm X-ray}-t_{\rm VLF}$ distributions for detected C- and M-class flares. $N$ is the number of events satisfying the operational detection criterion for the corresponding VLF--X-ray pairing. $N_{\rm VLF\,later}$, $N_{\rm co}$, and $N_{\rm VLF\,earlier}$ denote events with $\Delta t_{\rm peak}<-20$~s, $|\Delta t_{\rm peak}|\leq20$~s, and $\Delta t_{\rm peak}>20$~s, respectively. $N_0$ gives the number of exact zero-delay cases and is a subset of $N_{\rm co}$. Positive $\Delta t_{\rm peak}$ indicates that the VLF response extremum precedes the X-ray peak, whereas negative values indicate that the X-ray peak occurs first.
}

\label{tab_dt_stats}
\end{table*}

\section{Discussion}
\label{discussion}

\subsection{Detection Dependence on Observation Frequency}

The systematically higher detection efficiency at 21.4~kHz compared to 24.0~kHz may be understood in terms of ionospheric reflection physics. Lower VLF frequencies are reflected at higher effective altitudes in the D-region, closer to the heights where flare-driven soft X-ray ionization produces the largest relative changes in electron density \citep{Wait1964, Thomson1993}. As a result, small flare-induced perturbations lead to larger changes in effective reflection height and waveguide phase
path at lower frequencies, enhancing the amplitude and temporal coherence of the VLF response. This sensitivity is further amplified by the modal structure of the Earth--ionosphere waveguide, for which phase and interference patterns respond more strongly to reflection-height variations at lower VLF frequencies \citep{Budden1961, Ferguson1998}. Observational studies have consistently reported larger and more coherent VLF perturbations at lower frequencies during solar flares, particularly for weaker events near the detection threshold \citep{McRae2004, Clilverd2009}. Together, these effects may contribute to the higher detectability at 21.4~kHz observed in our statistical results.

The comparison between the two monitored frequencies indicates that 21.4~kHz provides the higher overall detection efficiency in the present dataset. Across the full sample, 21.4~kHz detects 70.8\% of all flares (177/250), compared with 62.0\% (155/250) at 24.0~kHz. The difference is also evident for M-class flares, for which 21.4~kHz detects 92.1\% (35/38) of events, while 24.0~kHz detects 76.3\% (29/38).
 
However, the use of two frequencies provides clear added value in terms of robustness and confidence. While 81.2\% of all flares (203/250) are detected in at least one VLF--X-ray pairing, 51.6\% (129/250) are independently confirmed at both frequencies. For M-class flares, dual-frequency confirmation is achieved for 73.7\% of all events (28/38) and for 77.8\% (28/36) of the detected subset, indicating that stronger flares frequently produce responses observable along both propagation
paths. The second frequency thus provides an additional confirmation channel and helps identify events whose response is observed consistently in both VLF channels. In practical terms, the 21.4~kHz channel yields the larger number of detections in the present dataset, whereas a dual-frequency configuration provides a more restrictive subset of events detected along both propagation paths. For applications requiring reliable large-scale flare monitoring rather than maximal event counts, the combined two-frequency approach therefore provides useful complementary information.

\subsection{Limitations and implications for flare detection using long-baseline VLF}

The present work is intended as an initial assessment of whether long-baseline VLF propagation can be used as a practical proxy for solar flare occurrence and timing. In that context, the VLF response should be interpreted as a path-integrated diagnostic of changes in the Earth-ionosphere waveguide rather than a localized measurement of D-region conditions. This has several direct implications for reliability as a flare detector.

First, detectability may be influenced by illumination geometry along the propagation path. While we classify events by local day/night at the receiving site, the trans-hemispheric paths to Antarctica traverse regions with substantially different solar zenith angles and may include terminator crossings. Consequently, two flares of similar GOES class can produce markedly different VLF signatures depending on whether a sufficient fraction of the propagation path is sunlit at the relevant times. This limitation is intrinsic to long-range VLF remote sensing and must be considered when comparing event-to-event responses and when designing automated detection logic.

Second, the VLF signal exhibits strong diurnal structure and variable background fluctuations that can obscure flare signatures, particularly for weaker events near the adopted selection threshold. Although detrending via superposed epoch analysis removes the mean daily variation, residual variability remains, and this variability can reduce the correlation with the flare X-ray profile even when a flare-related perturbation is present. From an operational perspective, this implies that detection thresholds based solely on instantaneous amplitude excursions may be unreliable, and that a combination of detrending and time-profile matching to X-ray proxies (or multi-frequency confirmation) is preferable.

Third, because the VLF response is sensitive to the state of the waveguide, instrumental and geophysical factors unrelated to flares can introduce apparent flare-like excursions or reduce true flare signatures. These include changes in transmitter power stability, local electromagnetic interference, and path-dependent propagation effects. In addition, uncertainties in the effective lower boundary conditions along the propagation path, including ground conductivity contrasts and coastal crossings, can modulate amplitude and phase in ways that are not straightforward to separate from flare forcing. These effects are expected to be more pronounced for very long paths spanning heterogeneous surface conditions, and they motivate caution when interpreting marginal detections.

Fourth, timing alignment between VLF and GOES soft X-ray channels can show near-zero or even negative best-fit lags when correlation-based matching is used. In the detector framing, such cases should not be over-interpreted as physically anomalous without additional constraints. Several practical mechanisms can produce this behavior, including differing rise/decay asymmetries between the VLF response and the broadband X-ray channel, the limited lag-search window and cadence, and the influence of background variability that shifts the correlation extremum. We distinguish explicitly between resolved interior correlation extrema, $|\Delta t_*|<360$~s, and boundary-limited solutions at $|\Delta t_*|=360$~s. The latter are not interpreted as physical lag measurements, since the true correlation extremum may lie outside the search interval.

Even for resolved interior extrema, the correlation lag should be regarded primarily as an empirical timing indicator rather than as a direct measurement of the physical ionospheric response time. Differing rise and decay asymmetries between the VLF response and the broadband X-ray channels, finite data cadence, and background variability can shift the correlation extremum. Consequently, near-zero or negative correlation lags should not be over-interpreted physically without additional observational or modelling constraints.

Finally, multi-frequency analysis provides a practical pathway to improve robustness. Events for which both frequencies exhibit consistent, high correlation with GOES and comparable timing are more credible as flare detections than those appearing in only one channel. Conversely, discrepancies between the two frequencies are informative about the limits of detection under particular propagation conditions and motivate future work incorporating phase information, additional transmitters, and multi-station observations. Overall, the results presented here demonstrate that Antarctic long-baseline VLF observations can detect a substantial subset of flares, but they also show that detection efficiency is conditional and must be quantified with explicit criteria and quality controls when VLF is used as a large-scale flare detector.

\subsection{Sensitivity to the correlation threshold}

To assess the sensitivity of the operational detection results to the adopted correlation threshold, we repeated the analysis using $\rho_0=0.5$, 0.6, 0.7, and 0.8. In each case, a VLF--X-ray pairing is classified as detected when the dominant signed correlation extremum within the $\pm180$~s association interval satisfies $|\rho_*|\geq\rho_0$. The resulting detection efficiencies are summarized in Table~\ref{tab:rho_sensitivity}.

As expected, the number of detections decreases as the required correlation strength is increased. For the full sample, the ``any-pairing'' detection efficiency decreases from 93.6\% (234/250) at $\rho_0=0.5$ to 88.0\% (220/250) at $\rho_0=0.6$, 81.2\% (203/250) at the adopted value $\rho_0=0.7$, and 63.6\% (159/250) at $\rho_0=0.8$. The dependence on threshold is considerably stronger for C-class than for M-class flares. The C-class detection efficiency decreases from 92.9\% (197/212) at $\rho_0=0.5$ to 58.5\% (124/212) at $\rho_0=0.8$. In contrast, the M-class detection efficiency remains high over the entire threshold range, changing only from 97.4\% (37/38) to 92.1\% (35/38). This difference indicates that the stronger M-class events generally produce VLF--X-ray correlations that remain above even the more restrictive thresholds, whereas a larger fraction of C-class detections lie closer to the adopted correlation cutoff.

The frequency-level results show a similar threshold dependence. At $\rho_0=0.5$, both VLF frequencies detect 217/250 events (86.8\%). For thresholds of 0.6 and above, the observed detection efficiency is higher at 21.4~kHz than at 24.0~kHz: 78.0\% versus 75.2\% at $\rho_0=0.6$, 70.8\% versus 62.0\% at $\rho_0=0.7$, and 53.6\% versus 46.0\% at $\rho_0=0.8$. As discussed in Sect.~4.1, these differences describe the performance of the complete transmitter--propagation-path--receiver channels and should not be interpreted as an isolated intrinsic frequency effect.

Requiring detection at both VLF frequencies is more restrictive at all thresholds. The dual-frequency detection fraction decreases from 80.0\% (200/250) at $\rho_0=0.5$ to 65.2\% (163/250) at $\rho_0=0.6$, 51.6\% (129/250) at $\rho_0=0.7$, and
36.0\% (90/250) at $\rho_0=0.8$. Thus, increasing the correlation threshold progressively selects a smaller subset of events with stronger and more consistently observed VLF--X-ray correspondence.

We retain $\rho_0=0.7$ as the operational threshold used throughout the main analysis. It provides a relatively restrictive requirement for VLF--X-ray correspondence while retaining 81.2\% of the full flare sample and 94.7\% of the M-class events. The threshold test also shows that the main sample-level result---the higher detectability of M-class than C-class flares---persists throughout the investigated range. Absolute detection efficiencies, particularly for C-class events and for dual-frequency confirmation, are nevertheless sensitive to the chosen value of $\rho_0$ and should therefore be interpreted with reference to the explicitly stated operational threshold.

\begin{table*}
\centering
\begin{tabular}{lcccc}
\toprule
Metric & $\rho_0 \geq 0.5$ & $\rho_0 \geq 0.6$ & $\rho_0 \geq 0.7$ & $\rho_0 \geq 0.8$ \\
\midrule
All flares (any pairing)
& 93.6 (234) & 88.0 (220) & 81.2 (203) & 63.6 (159) \\

C-class (any pairing)
& 92.9 (197) & 86.8 (184) & 78.8 (167) & 58.5 (124) \\

M-class (any pairing)
& 97.4 (37) & 94.7 (36) & 94.7 (36) & 92.1 (35) \\

21.4~kHz (all flares)
& 86.8 (217) & 78.0 (195) & 70.8 (177) & 53.6 (134) \\

24.0~kHz (all flares)
& 86.8 (217) & 75.2 (188) & 62.0 (155) & 46.0 (115) \\

Both frequencies
& 80.0 (200) & 65.2 (163) & 51.6 (129) & 36.0 (90) \\
\bottomrule
\end{tabular}

\caption{
Detection efficiency (\%) as a function of the adopted correlation threshold $\rho_0$. For each VLF--X-ray pairing, the dominant signed correlation extremum $\rho_*$ is determined within the operational $\pm180$~s association window, and the pairing is classified as detected when $|\rho_*|\geq\rho_0$. ``Any pairing'' denotes detection in at least one of the four VLF--X-ray pairings. The 21.4 and 24.0~kHz rows denote detection at the corresponding VLF frequency in at least one GOES XRS
channel, while ``Both frequencies'' requires detection at both VLF frequencies. The number of detected flares is shown in parentheses.}
\label{tab:rho_sensitivity}
\end{table*}

\section{Conclusions}
\label{conclusions}

The results presented here demonstrate that long-baseline VLF observations from Antarctica provide a viable remote-sensing proxy for solar flare occurrence and timing. Using continuous monitoring of two trans-hemispheric VLF transmitters at 21.4 and 24.0~kHz, we analysed the ionospheric response to 250 C- and M-class solar flares observed by GOES during the period 24 January--8 February 2025. After removing the strong diurnal variation of the waveguide signal, flare signatures were identified through cross-correlation between VLF amplitude perturbations and the GOES soft X-ray flux. Under this operational definition, 81.2\% (203/250) of the flares in the sample exhibit detectable VLF responses in at least one VLF--X-ray pairing. The detection efficiency is 78.8\% (167/212) for C-class flares and 94.7\% (36/38) for M-class events. At the frequency level, 70.8\% (177/250) of the flares are detected at 21.4~kHz and 62.0\% (155/250) at 24.0~kHz, while 51.6\% (129/250) are
detected at both VLF frequencies.

The analysis shows that the lower-frequency path (21.4 kHz) consistently yields equal or higher detection efficiency than 24.0 kHz. Multi-frequency observations provide additional robustness, allowing independent confirmation of flare signatures and improving confidence in detections. The results also demonstrate that detection efficiency depends not only on flare class but is likely influenced by the illumination conditions along the propagation path, emphasizing the importance of path-integrated ionospheric conditions for interpreting long-baseline VLF measurements.

Overall, the Antarctic VLF observations presented here confirm that long propagation paths can provide enhanced sensitivity to flare-driven ionospheric disturbances and can detect a substantial fraction of GOES-class events. At the same time, the analysis highlights intrinsic limitations arising from waveguide variability, mixed illumination along the path, and background fluctuations. Future work combining multi-frequency VLF measurements with phase observations, additional transmitters, and spectral X-ray data will further improve the reliability of VLF-based flare monitoring and help to better constrain the response of the ionospheric D-region to solar eruptive activity.

\section*{Acknowledgments}
This study was funded by the Bulgarian Ministry of Education and Science through the National Centre for Polar Studies, and Sofia University “St. Kliment Ohridski” in the framework of the National Program for Polar Studies 2022–2025, grant number D70-25-44/12.04.2024. We acknowledge funding from the LOFAR-BG project of the National Roadmap for Research Infrastructure of Bulgaria, under contracts D01-362/14.12.2023 and D01-110/30.06.2025 with the Ministry of Education and Science. Part of the instrumentation was developed with financial support from NextGenerationEU, under the National Recovery and Resilience Plan of the Republic of Bulgaria (Project No. BG-RRP-2.004-0005)

\section*{Annex I}
\onecolumn
\begin{longtable}{rcccccccc}
\caption[]{List of the solar flares, which occurred during the period of observations. Information from NOAA/SWPC.}\\
\label{flare_list}\\
\toprule
ID & Start & Peak & End & Lon & Lat & AR & Class & Is Day \\
\midrule
\endfirsthead

\toprule
ID & Start & Peak & End & Lon & Lat & AR & Class & Is Day \\
\midrule
\endhead
\midrule
\multicolumn{9}{r}{Continued on next page} \\
\midrule
\endfoot
\bottomrule
\endlastfoot
1 & 2025-01-24 00:28 & 00:33 & 00:51 & -20 & 5 & 13972 & C3.5 & Yes \\
2 & 2025-01-24 04:02 & 04:10 & 04:19 & 12 & 36 & 13971 & C4.4 & No \\
3 & 2025-01-24 20:48 & 21:04 & 21:17 & -6 & 67 & 13961 & M2.7 & Yes \\
4 & 2025-01-25 00:18 & 00:22 & 00:26 & -10 & 67 & 13961 & C6.6 & Yes \\
5 & 2025-01-25 02:52 & 03:00 & 03:18 & -11 & 66 & 13961 & C3.7 & No \\
6 & 2025-01-25 03:16 & 03:23 & 03:28 & -11 & 69 & 13961 & C4.7 & No \\
7 & 2025-01-25 03:41 & 03:47 & 03:51 & 18 & 87 & 13959 & C3.4 & No \\
8 & 2025-01-25 04:05 & 04:35 & 04:52 & -10 & 67 & 13961 & C3.3 & No \\
\end{longtable}
\twocolumn

\section*{Annex II}
\onecolumn

\begin{longtable}{lrrrrrrrrrrrr}
\caption{Signed VLF--GOES cross-correlation results over the $\pm360$~s lag-search interval. For each VLF--X-ray pairing, $\rho_{\max}$ and $\rho_{\min}$ are the maximum and minimum values of the cross-correlation function, respectively. The coefficient with the larger absolute magnitude, and therefore retained as the dominant signed correlation extremum, is shown in bold; the corresponding lag is reported as $\Delta t_{*}$. Boundary-limited solutions with $|\Delta t_{*}|=360$~s are marked with $\dagger$ and are retained for completeness but are not interpreted as resolved lag measurements.}\label{tab_correlation}\\

\toprule
\multicolumn{1}{c}{} &
\multicolumn{3}{c}{\textbf{21.4~kHz -- X-ray A chan.}} &
\multicolumn{3}{c}{\textbf{21.4~kHz -- X-ray B chan.}} &
\multicolumn{3}{c}{\textbf{24.0~kHz -- X-ray A chan.}} &
\multicolumn{3}{c}{\textbf{24.0~kHz -- X-ray B chan.}} \\

ID
& $\Delta t_{*}$ & $\rho_{\max}$ & $\rho_{\min}$
& $\Delta t_{*}$ & $\rho_{\max}$ & $\rho_{\min}$
& $\Delta t_{*}$ & $\rho_{\max}$ & $\rho_{\min}$
& $\Delta t_{*}$ & $\rho_{\max}$ & $\rho_{\min}$ \\

\midrule
\endfirsthead

\toprule
\multicolumn{1}{c}{} &
\multicolumn{3}{c}{\textbf{21.4~kHz -- X-ray A chan.}} &
\multicolumn{3}{c}{\textbf{21.4~kHz -- X-ray B chan.}} &
\multicolumn{3}{c}{\textbf{24.0~kHz -- X-ray A chan.}} &
\multicolumn{3}{c}{\textbf{24.0~kHz -- X-ray B chan.}} \\

ID
& $\Delta t_{*}$ & $\rho_{\max}$ & $\rho_{\min}$
& $\Delta t_{*}$ & $\rho_{\max}$ & $\rho_{\min}$
& $\Delta t_{*}$ & $\rho_{\max}$ & $\rho_{\min}$
& $\Delta t_{*}$ & $\rho_{\max}$ & $\rho_{\min}$ \\

\midrule
\endhead

\midrule
\multicolumn{13}{r}{Continued on next page} \\
\midrule
\endfoot

\bottomrule
\endlastfoot

1 & -30 & -0.37 & \textbf{-0.63} & -30 & -0.44 & \textbf{-0.65} & -320 & -0.60 & \textbf{-0.89} & -320 & -0.67 & \textbf{-0.89} \\
2 & 220 & 0.47 & \textbf{-0.78} & 200 & 0.34 & \textbf{-0.91} & 220 & 0.39 & \textbf{-0.78} & 220 & 0.15 & \textbf{-0.86} \\
3 & 160 & \textbf{0.92} & 0.06 & 90 & \textbf{0.97} & 0.40 & 150 & \textbf{0.87} & 0.14 & 70 & \textbf{0.94} & 0.46 \\
4 & -340 & 0.87 & \textbf{-0.95} & -340 & 0.89 & \textbf{-0.96} & -330 & 0.87 & \textbf{-0.95} & -330 & 0.86 & \textbf{-0.95} \\
5 & 100 & \textbf{0.72} & -0.51 & 90 & \textbf{0.74} & -0.53 & 120 & \textbf{0.49} & -0.37 & 100 & \textbf{0.48} & -0.36 \\
6 & -170 & \textbf{0.76} & -0.68 & -210 & \textbf{0.71} & -0.69 & 360$^{\dagger}$ & \textbf{0.56} & -0.33 & -160 & \textbf{0.62} & -0.26 \\
7 & 360$^{\dagger}$ & 0.40 & \textbf{-0.64} & 360$^{\dagger}$ & 0.31 & \textbf{-0.61} & 360$^{\dagger}$ & 0.17 & \textbf{-0.67} & 240 & 0.14 & \textbf{-0.65} \\
8 & 360$^{\dagger}$ & 0.35 & \textbf{-0.53} & 360$^{\dagger}$ & -0.35 & \textbf{-0.90} & 360$^{\dagger}$ & \textbf{0.34} & -0.31 & 360$^{\dagger}$ & \textbf{0.74} & 0.29 \\
\end{longtable}
\twocolumn
%\bibliographystyle{cas-model2-names}
%\bibliography{bibliography_vlf}

\end{document}